\documentclass[graybox]{svmult}

\usepackage{mathptmx}       
\usepackage{helvet}         
\usepackage{courier}        
\usepackage{type1cm}        
\usepackage{makeidx}         
\usepackage{graphicx}        
\usepackage{multicol}        
\usepackage[bottom]{footmisc}

\makeindex             

\begin{document}

\title*{Photometric and Kinematic Characterization of Tidal Dwarf Galaxy candidates}
\author{D. Miralles-Caballero$^1$, L. Colina$^1$ and S. Arribas$^1$}
\institute{D.Miralles-Caballero, L.Colina \& S.Arribas \at DAMIR-IEM-CSIC, Serrano 121 28006 Madrid, \email{dmiralles@damir.iem.csic.es}} 
%
%
\maketitle

\abstract*{Tidal Dwarf Galaxies (TDG), or self-graviting objects
 created from the tidal forces in interacting galaxies, have been found in several
  merging systems. This work will focus on identifying TDG candidates among a sample of 
  Luminous and Ultraluminous Infrared Galaxies (U)LIRGs, where these interactions are
  occurring, in order to study their formation and evolution. High angular resolution imaging
   from Hubble Space Telescope (HST) in B, I and H band will be used to detect these sources.
   Photometric measurements of these regions compared to Stellar Synthesis Population
    models will allow us to roughly estimate the age and the mass. Complementary optical
     Integral Field Spectroscopy we will be 
     able to explore the physical, kinematic and dynamical properties in TDGs. We present 
     preliminary photometric results for IRAS 0857+3915, as an example of the study that will
      be held for the entire sample of (U)LIRGs.}

\abstract{Tidal Dwarf Galaxies (TDG), or self-graviting objects
 created from the tidal forces in interacting galaxies, have been found in several
  merging systems. This work will focus on identifying TDG candidates among a sample of 
  Luminous and Ultraluminous Infrared Galaxies (U)LIRGs, where these interactions are
  occurring, in order to study their formation and evolution. High angular resolution imaging
   from Hubble Space Telescope (HST) in B, I and H band will be used to detect these sources.
   Photometric measurements of these regions compared to Stellar Synthesis Population
    models will allow us to roughly estimate the age and the mass. Complementary optical
     Integral Field Spectroscopy we will be 
     able to explore the physical, kinematic and dynamical properties in TDGs. We present 
     preliminary photometric results for IRAS 0857+3915, as an example of the study that will
      be held for the entire sample of (U)LIRGs.}
      
\section{Introduction}
\label{sec:1}

A large fraction of Luminous Infrared Galaxies, LIRGs (L$_{IR}$ = L(8-1000µm) = 10$^{11}$ - 10$^{12}$ L$\odot$), 
and most Ultraluminous Infrared Galaxies, ULIRGs (L$_{IR}$ $>$ 10$^{12}$ L$\odot$; 
see~\cite{sanders96}), show signs of mergers and interactions (e.g.~\cite{veilleux02}). 
According to models (eg. ~\cite{duc04}), knots of  star formation outside the nuclei, with masses $\sim$ 10$^8$ - 10$^9$ M$\odot$, 
can be formed from the debris of the interaction : the Tidal Dwarf Galaxies, TDGs. Besides, many
knots associated with star formation have already been observed at the outskirts in other merging 
systems (eg.~\cite{duc98}). (U)LIRGs are the ideal laboratory to study TDG candidates, since
they represent one of the most extreme cases of galaxy merging.

Our main goal will be a systematic search for TDG candidates and their physical, 
kinematic and dynamical study among 34 low z (U)LIRGs. We report on the preliminary 
photometric results for the ULIRG IRAS 0857+3915 (L$_{IR}$ = 12.15 L$\odot$), which will exemplify
 the photometry we are carrying out for the complete sample.

\begin{figure}[!hb]
\sidecaption
\includegraphics[scale=.47]{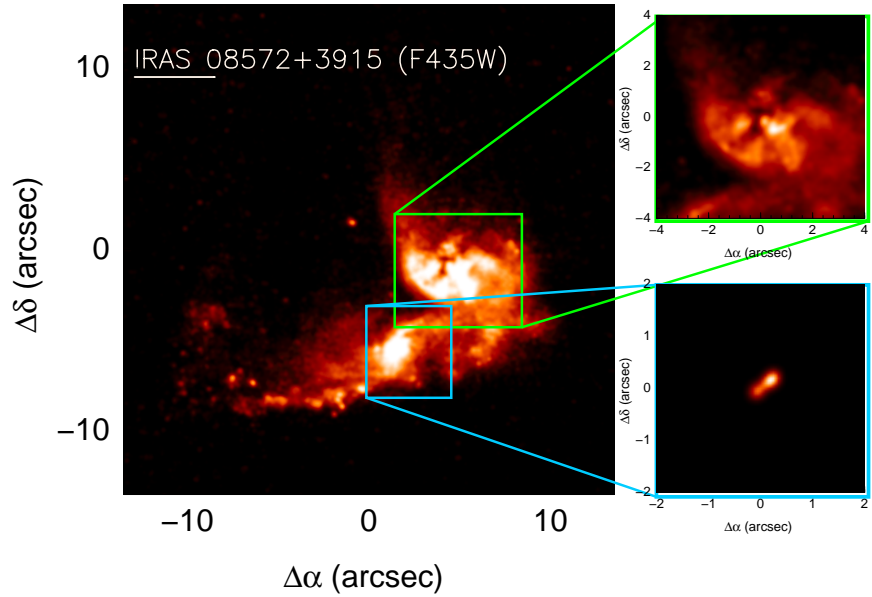}
%
%
\caption{IRAS 0857+3915 consists of 2 spiral galaxies that have not merged yet.
 As it can be seen, a tail extends up to 30 kpc from the East nucleus, bending to
  North-West direction. There is another tail  going up straight North from the West
   nucleus. Throughout the tails many blue knots that could be associated with star
   formation regions can be observed.}
\label{fig:1}       
\end{figure}

\section{Sample}
\label{sec:2}

Our sample includes 20 LIRGs and 14 ULIRGs that have been selected from 
the HST database available with at least blue F435W ($\sim$ B) and red F814W 
($\sim$ I) filters with the camera ACS. These systems present different morphologies, 
that covers all merging phases, and cover distances up to ~ 400 Mpc. We also selected
H band images (filter F160W) with the same camera when possible, useful to study 
the inner regions of the sources.

\section{IRAS 08752+3915 Photometry: identifying TDG candidates}
\label{sec:3}
\subsection{Photometric measurements}
\label{subsec:2}


%
We selected our possible TDG candidates for our photometry as the bright condensations above 3 
sigma of the local background level, lying outside de nuclear region 
(i.e., $>$ 2kpc) in our images (see figure 1). \\

All photometric calibrations and magnitude determinations of these regions were performed
 following the prescriptions outlined in~\cite{sirianni05}. Most of the magnitudes were derived from aperture photometry 
 from circular apertures typically 4-7 pixels (scale of 0.05 ''/pixel)in radius, although 
 in some cases polygonal apertures were needed due to the irregular shape of the regions.
 We measured the flux for the total region and also estimated the underlying galaxy flux (tail flux) 
 by using the mean of the pixels in a 5-pixel annulus starting 8 pixels away from the center 
 of the condensation, more than 8 times  the width of the PSF but close engouh to really 
 estimate its nearby local background. The knot flux is defined as the total region flux 
 subtracting the tail flux. These  measurements will allow us to study the relative
  ages and masses of both knots and the underlying galaxy in each case. Aperture of 1.5'' 
  was centered at the nuclei and no correction for internal extinction in the parent
  galaxy has been applied. \\

\begin{figure}[!ht]
\sidecaption
\includegraphics[scale=.70]{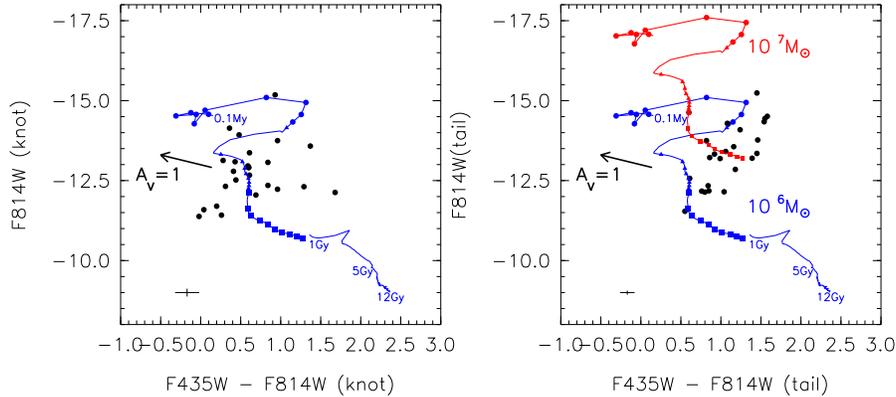}
%
%
\caption{Comparison between the model chosen and absolute magnitudes of the knots (left)
 and the tails (right). The blue curve is normalized to 10$^6$M$\odot$ and the red one
  to 10$^7$M$\odot$. Typical error for regions  is plotted on the bottom left side corner.
  Arrows indicating the
 effects of 0.5 mag deredening in V are also plotted. Ages are indicated such a way that 
 circles stand for steps of 1 Myr, triangles of 10 Myr and squares of 100 Myr}
\label{fig:2}       
\end{figure}

Photometric measurements will be compared to a stellar population 
synthesis model to roughly derive the age and mass of the regions.
Starburst99 (SB99,~\cite{leitherer99}) are used, since they are suitable for
young population. We will compare our data to an instantaneous burst model with solar
metalicity  and Salpeter IMF (same as previous studies) with lower mass limit of 
0.1 M$\odot$ (with 1M $\odot$ would scale the derived mass to 0.4) and  upper mass limit
of 120 M$\odot$. We chose this to be instantaneous because it is the type of burst expected 
for young population.
\subsection{Properties of the TDG candidates}
\label{subsec:2}

Figure~\ref{fig:2} shows the preliminary results of the comparison between the model 
and the data of the 25 regions found in IRAS 08752+3915. From it some properties for 
the possible TDG candidates found can be inferred:
\begin{enumerate}
\item{Average magnitude differences between tail and knots in I band
 suggests that most of the mass in the tail corresponds to old population, which is what we would 
 expect.} 

\item{Around 70 \% of the knots have a B-I color value below 0.7. Hence, their
estimated age cannot be more than 300 Myr. Some of them  could even be less 
than 10 Myr, dependening on the extintion and mass degeneration. 
Assuming this, total region (knot + underlying galaxy) stellar mass is estimated 
to be around 10$^6$-10$^7$ M$\odot$ for most of the regions.}
\end{enumerate}

The forhtcoming spectral study will help us further characterize the properties of 
 these TDG candidates and, therefore,  better understand their evolution.

\begin{acknowledgement}
This work has been supported by the Spanish Ministry for Education and Science under
grant ESP2007-65475-C02-01.
\end{acknowledgement}

\end{document}